\documentstyle[prb,aps,epsfig,twocolumn]{revtex}

\draft
\title{\bf Unusual magneto-elastic forces stabilize the insulating ferromagnetic state of manganites: the La$_{0.8}$Ca$_{0.2}$MnO$_3$ case.
}
\author{ M.Hennion$^1$, F. Moussa$^1$, F. Wang, J. Rodr\'{\i}guez-Carvajal$^1$, Y. M. Mukovskii$^2$ and D. Shulyatev$^2$}

\vspace{0.5cm}

\address{$^1$Laboratoire L\'eon Brillouin, CEA-CNRS, CE Saclay, 
91191 Gif sur Yvette Cedex, France}
\address{$^2$Moscou State Steel and Alloys Institute, Leninskii prospect 4, Moscou 117936 Russia} 
\date{\today, submitted to Phys. Rev. B}
\begin{document}
\twocolumn[\hsize\textwidth\columnwidth\hsize\csname @twocolumnfalse\endcsname
\maketitle
\begin{abstract}
The ferromagnetic and insulating state observed in La$_{1-x}$Ca$_{x}$MnO$_3$, 0.125$<$x$<$0.2, is characterized by structural and magnetic anomalies below T$_C$,  similar to those observed in the x$_{Sr}$$\approx$1/8. A neutron scattering study of the superlattice {\bf Q$_0$}= (0,0,1+/4)$_{cub}$ peak, and of the magnetic excitations are reported in the x$_{Ca}$=0.2 sample. The occurrence of this superstructure is associated with the observation of a gap in the spin dynamics, at a {\bf q$_0$} wave-vector ({\bf q$_0$}={\bf Q$_0$}-$\tau$) with the same modulus $\vert${\bf q$_0$}$\vert$ in all directions, which divides the dispersion into two regimes. For $\vert${\bf q}$\vert$$<$$\vert${\bf q$_0$}$\vert$ the dispersion is splitted into two or three curves. For $\vert${\bf q}$\vert$$>$$\vert${\bf q$_0$}$\vert$, magnetic excitations lock on acoustic and optic phonon energies, revealing a new kind of magneto-vibrational coupling. 
 We suggest an analysis in terms of two distinct magnetic couplings, associated with two ferromagnetic media involved into a collective state.
\end{abstract}
]
The existence of a ferromagnetic insulating state, occurring at a temperature 
T$_{O'O"}$$\approx$150K
 lower than the ferromagnetic transition T$_C$
%$\approx$190K, 
in  La$_{1-x}$Sr$_x$MnO$_3$ (0.1$<$x$<$0.17), is one of the most puzzling observation in manganites. Structural changes, with new superlattice Bragg peaks, ({\it h,h,2l+1})$_{ortho}$ and ({\it h,h,l+1/2})$_{ortho}$, are observed\cite{Yamada,Yamada2}, 
 whereas a decrease of the orthorhombicity occurs, associated with Mn-O lengths becoming nearly equal\cite{Pinsard}.
At the same temperature a small jump of the magnetization\cite{Endoh} occurs. This metal-insulator transition, has been explained by a polaron ordering\cite{Yamada,Yamada2} or a new type of orbital ordering\cite{Endoh}, but a definitive interpretation is still lacking.
The same transport-magneto-structural anomalies are observed in La$_{1-x}$Ca$_x$MnO$_3$ (0.125$\le$x$\le$0.20). An upturn of the resistivity has been reported at a temperature below T$_C$\cite{Okuda}.
The othorhombicity characteristic of the staggered orbital ordering which occurs at T$_{JT}$ {\it above T$_C$}, decreases below T$_C$, at T$\approx$100-80K\cite{Biotteau}.  An anomalous increase  of intensity in some ferromagnetic Bragg peaks has also been observed defining a line, T$_B$(x), in the phase diagram, with T$_B$$\approx$80K for x$_{Ca}$=0.2. There are some differences between the two systems. For instance, the ({\it h,h,2l+1})$_{ortho}$ peaks are observed in the Ca system at any temperature with a very small intensity, instead of below T$_{O'O"}$ for the Sr doped case. The presence of ({\it h,h,l+1/2})$_{ortho}$ peaks have not been reported up to now in the Ca system. Moreover, this transition is sharply defined at T$_{O'O"}$ in the x$_{Sr}$=0.125 case, whereas it is smooth in the Ca-doped case, suggesting some inhomogeneous character. 

 In the present paper, devoted to the x$_{Ca}$=0.2 compound (T$_C$=180K), new measurements of the  superlattice  ({\it 0,0,1+1/4})$_{cub}$ peak  or ({\it 0,0,1+1/2}$_{ortho}$) and of magnetic and  lattice excitations are reported, which shed a new light on the low temperature state. 
We observe a direct correspondence in temperature between this new periodic structure and the occurrence of an energy gap in the spin dynamics in all directions, {at a q wave-vector with 
the same modulus {\bf $\vert$q$_0$$\vert$}.   This wave vector divides the dispersion curve into two regimes,  {\bf $\vert$q$\vert$$<$$\vert$q$_0$$\vert$} where the curve is splitted into two or possibly three curves and {\bf $\vert$q$\vert$$\ge$$\vert$q$_0$$\vert$} where magnetic excitations lock at acoustical and optical phonon modes, revealing a new kind of magneto-vibrational coupling. We suggest an analysis in terms of two distinct magnetic couplings, associated with two ferromagnetic media, involved into a collective state.

Experiments have been carried out at the reactor  Orph\'ee (Laboratoire L\'eon Brillouin) on triple axis spectrometers, set at the thermal and cold source. For simplicity we 
use the cubic indexation. As usual, we define {\bf Q}={\bf q}+$\tau$. The main results are summarized now.

At T$\approx$90-100K, below T$_C$=180K, the superlattice (0,0,1/4)$_{cub}$ peak intensity (or (0,0,1/2)$_{ortho}$), also observed in the x=1/8 Sr-doped compound, appears with a small intensity, superposed over a large background I$_B$, characteristic of a magnetic and atomic disorder. This is shown in Fig. 1-a. 
This peak is actually metastable since it may not develop depending on the cooling rate. It defines a new 4a$_0$ periodicity (a$_0$ is the cubic edge) with a large correlation length $\xi$ ($\approx$300$\AA$). Below T$_f$=45K, an increase of the q-linewidth indicates a slight decrease of the correlation length (Fig 1-b). This last feature, absent for x$_{Sr}$=1/8, reveals some frustration effects. Along [001], I$_B$ reported as cooling down, exhibits an interesting temperature variation.  
A sharp decrease is observed below T$\approx$75$-$80K, concomitantly with the step-increase of ferromagnetic Bragg peaks. A further increase of I$_B$ is observed below 45K,
as the correlation length of the superstructure decreases. This variation also shows irreversibility.
%mimics that of the magnetization $m(T)$ obtained under applied field, which probes the pinning of %the magnetic domains performed in this sample.

This transition at T$\approx$90-100K, characteristic of the propagation vector {\bf q$_0$}=(0,0,1/4), is related to the occurrence of an isotropic energy "gap" at $\vert${\bf q$_0$}$\vert$ in the spin wave dispersion defining two regimes, {\bf q$<$q$_0$} and {\bf q$\ge$q$_0$}.  We first describe the magnetic excitation spectrum in the low temperature state (T=17K), along [001], shown in Fig. 2-(a), (d) and (e). In the smallest q range, (Fig 2-(d)), two, or possibly three modes -when using appropriate resolution- are observed, showing a similar quadratic law, and distinct anisotropy gaps. An example of energy spectrum is seen in the inset of Fig. 2-(d). As {\bf q} increases, in the intermediate regime reported in Fig 2-(e), a slight departure from the quadratic law is observed for the upper mode which merge, at {\bf q$_0$}, into an energy close to the longitudinal acoustic phonon energy (LA). The dispersion curve of the lower energy remains quadratic nearly up to {\bf q$_0$} where it bends at the TA value. Its intensity vanishes beyond {\bf q$_0$}, defining an energy gap. For {\bf q}$\ge${\bf q$_0$} the magnetic dispersion curve is broken into several parts. As seen in Fig. 2-a where phonon dispersions are shown by dotted lines, these discontinuities may reflect the locking of magnetic excitations on phonons as q increases up to the zone boundary: from the TA phonon branch to modes close to the LA phonon branch at {\bf q$_0$} and beyond, and then to modes close to the longitudinal optical phonon branch (LO). 
\begin{figure}
\centerline{\epsfig{file=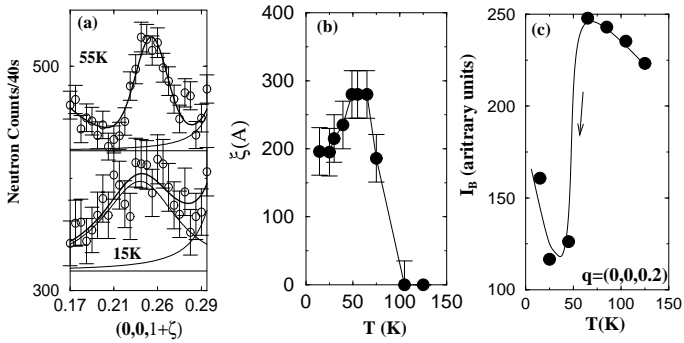,width=9cm}}
\caption{ {\bf a}: typical q-dependent spectrum of the superlattice peak measured at {\bf Q$_0$}=(0,0,1.25) at two temperatures, fitted by a gaussian function (full line). {\bf b}: temperature variation of the correlation length of the superlattice peak. {\bf c}: temperature variation of the incoherent background measured at {\bf Q}=(0,0,1.2) by cooling. In {\bf b, c}, the full line is a guide for the eyes.}
\end{figure}

Very similar anomalies 
 have been observed below T$_{O'O"}$ in the spin dynamics of the x$_{Sr}$=0.125 compound\cite{Moussa}. In that case, the well-defined peaks, likely associated with the well-developed superstructure, allow to clearly identify an intermediate regime, {\bf q$_0$/2}$<${\bf q}$<${\bf q$_0$}, where a deviation from the quadratic law can be seen for the upper mode in Fig. 2-(f).

 A gap at {\bf q$_0$} has also been found at low temperature
along [001] in the x$_{Sr}$=0.15\cite{Doloc}. 

We conclude that, {\it this unusual spin dynamics is typical of the ferromagnetic insulating state}.

The unusual character of the magnetic excitations can be also observed along [110] and [111], although with experimental difficulties due to very broad modes. This is shown in Fig. 2-(b), (c). The most remarkable observation is that the gap anomaly found at {\bf q$_0$} along [001] is observed at {\bf q'$_0$} and {\bf q''$_0$} in the other symmetry directions, with the same modulus ( $\zeta$=0.25, 0.18 and 0.14 in reduced lattice units along [001], [110] and [111], in Fig. 2-(a),(b),(c) respectively). 

\begin{figure}
\centerline{\epsfig{file=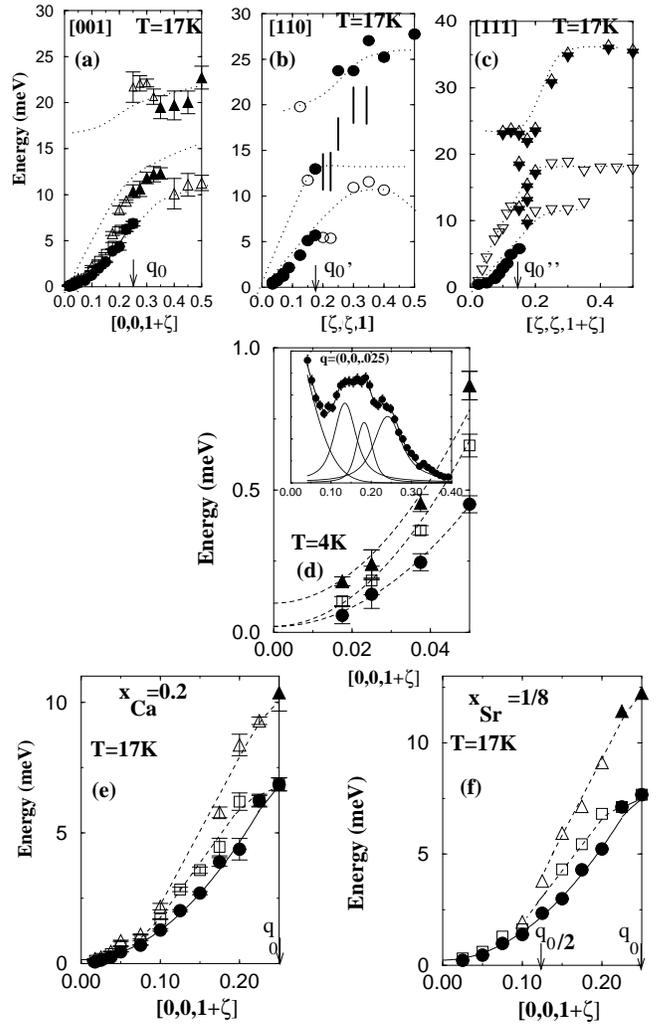,width=8.4cm}}
\caption{{\bf (a), (b), (c)}: spin wave dispersions for q along  [001]$_{cub}$, [110]$_{cub}$ and [111]$_{cub}$
in reduced lattice units ({\it rlu}). The dotted lines are phonon dispersions. The arrows locate the gap at {\bf q$_0$}, {\bf q'$_0$}, and {\bf q''$_0$}, showing the same modulus. In {\bf (a), (b)}, the filled (empty) symbols are excitations with the larger (weaker) intensity. In {\bf (c)},
the filled, half-filled and empty symbols refer to purely magnons, mixed modes and purely phonons. In {\bf (b)}, the vertical lines indicate very broad modes also observed in some cases (irreversibilities). {\bf (d)}: magnetic excitations in the smallest q range, fitted with quadratic laws. Inset: energy spectrum at q=0.025 in {\it rlu} along [001], with a very high resolution ( k$_f$=1.05 $\AA^{-1}$). {\bf (e), (f)}:
magnetic excitations in the $\vert${\bf q}$\vert$$<$$\vert${\bf q$_0$}$\vert$ range along [001] for x$_{Ca}$=0.2 ({\bf (d)}) and for x$_{Sr}$=1/8 ({\bf (e)}). In both cases, the filled (empty) symbols are excitations with the larger (weaker) intensity, the continuous line is a quadratic law, and the dashed one, a guide for the eye.}
\end{figure} 

The interest of this x$_{Ca}$=0.2 compound comes from the fact that the direct correspondence between the superstructure (0,0,1/4)$_{cub}$ and the spin dynamics can be seen both at 90K where the superstructure occurs and below 45K where its correlation length stops of increasing and slightly decreases. 

 In Fig 3-(a), energy spectra observed along [001] at {\bf q$_0$} are reported for several temperatures. The LA and TA phonon energies, temperature -independent, determined in farther Brillouin zones are indicated by vertical lines. At 125K, the magnetic energy spectrum consists of a very broad mode centered at the TA energy, which renormalizes from temperatures higher than T$_C$.
% A residual quasi-elastic signal, indicates some disordered spins. 
As T decreases below 100K, some  magnetic intensity rises on the upper energy side of the broad mode, corresponding to the LA energy. At lower temperature, a slight softening of this mode is observed, so that
 at 17K, in Fig 2-(a), the magnetic dispersion curve lies slightly below the phonon dispersion one (dotted line). The temperature dependence of the two magnetic modes determined at {\bf q$_0$} is reported in Fig 4-(a) ((1) and (2)) with their intensity variation in Fig 4-(b). This latter variation shows that the occurrence of the gap in the dispersion curve below 90K corresponds mainly to a transfer of intensity from the lower mode to the upper one. This evolution is stopped below 45K. 
\begin{figure}
\centerline{\epsfig{file=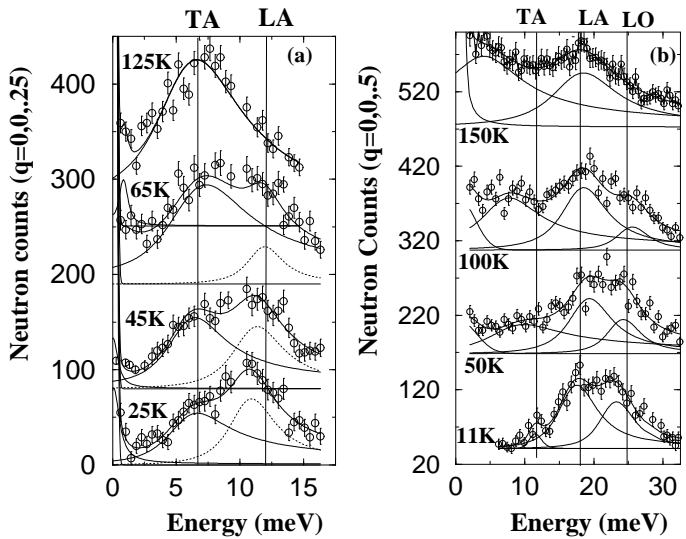,width=9cm}}
\caption{{\bf (a), (b)}: Energy spectra versus temperature at {\bf Q$_0$}=(0,0,1.25) (left panel) 
and {\bf Q}= (0,0,1.5) (right panel). The vertical bars locate the acoustical (TA and LA) and first optical (LO) phonon energies at the corresponding {\bf q}.}
\end{figure} 
 As another example, energy spectra obtained at the zone boundary {\bf Q}=(0,0,1.5) are shown in Fig 3-{\bf b}. They reveal a progressive transfer of magnetic intensity, as T decreases, from the magnetic low- energy mode which appears close to T$_C$ towards a high energy mode located at the LO phonon energy. This evolution starts at some temperature higher than the transition, which could be an indication for an inhomogeneous transition. A slight softening of this upper mode can also be observed below 50K. At 15K, a residual magnetic intensity at the TA energy is suggested (reported by empty triangles in Fig 2-{\bf a}). The broad energy mode located around 4Thz, actually persists up to 300K, indicating a purely phononic origin (LA), and, therefore, not reported in Fig. 2-{\bf a}. Finally, we mention that, at 300K, well-above T$_C$, large ferromagnetic fluctuations persist.

 The isotropic character of the gap anomaly is also seen in the temperature 
study. This has been specially studied at {\bf q'$_0$} along [110]. 
 Examples of energy spectra at two temperatures are reported in Fig. 4-(c). In Fig. 4-(b), the intensity variation of the two magnetic excitations located at TA and LA reveals that the upper energy mode (LA) appears below 90K with the superstructure and that this variation is stopped below 45K, as observed at {\bf q$_0$}.

\begin{figure}
\centerline{\epsfig{file=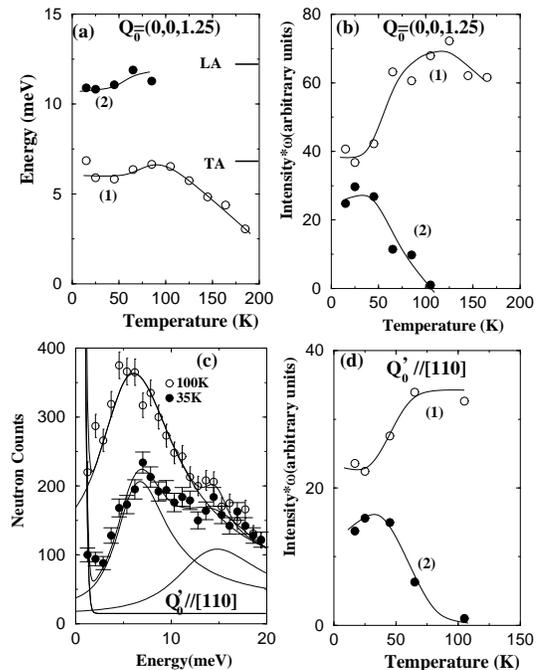,width=7cm}}
\caption{{\bf a}: Temperature variation of the magnetic energy modes at {\bf Q$_0$}=(0,0,1.25). The LA and TA energies are indicated on the right scale. {\bf b}: Temperature intensity variation of  the upper (filled circle) and the lower (empty circle) energy modes at {\bf Q$_0$}. {\bf c}: Exemple of spectra at {\bf Q'$_0$}//[110], at two temperatures. The residual peak at 15meV at 100K is actually a spurious effect, since observed at all temperatures. {\bf d}: Temperature intensity variation of the upper (filled circle) and the lower (empty circle) energy modes measured at {\bf Q'$_0$}.}
\end{figure}

This complex spin dynamics can be analysed in terms of two magnetic couplings with distinct anisotropy gaps, associated with two ferromagnetic media, as discussed now.

 The observations at {\bf q}$>${\bf q$_0$}, which probe a scale smaller than 4a$_0$, reveal a new kind of coupling. Because of their magnetic character, we call them "magneto-vibrational" modes (MV) below. They differ, however, from usual MV modes, which exist at phonon energies distinct from magnon ones\cite{Lovesey}. The present observation reveals {\it a resonance of magnon energies with phonon ones}, more or less accurate, or mixed magnon-phonon excitations. The intensity of these excitations is modulated with q, so that it spreads over a large energy range which reflects approximately a ferromagnetic dispersion law, revealing the underlying r\^ole of a ferromagnetic coupling.  
 The observations at {\bf q}$<${\bf q$_0$} range, which probe larger scales, showing a splitting of the q-dispersion, reveal the coexistence of two spin couplings, of distinct origin. One, associated with the lowest energy, corresponds to a ferromagnetic coupling, likely of super-exchange (SE) and double-exchange (DE) types, with the peculiarity that it is not detected beyond {\bf q$_0$}. It determines a stiffness constant value, D=48meV$\AA^2$, which is the same as found for x$_{Ca}$=0.125\cite{Biotteau}.The other one,
associated with the upper excitation can be related to the new kind of coupling (MV type) described above for {\bf q$>$q$_0$}. This latter excitation evolves progressively from a mixed magnon-phonon to a usual ferromagnetic one, when considering a smaller-q scale (large spatial scale), where the corresponding dispersion curve is found quadratic. In the same way, the dispersion curve at intermediate energies
 could be related to the new MV coupling, associated with the transverse phonon branch.
%This evolution with q is better observed in the case of the x=1/8 (Fig. 2-f). 

These observations can be interestingly compared with those found in the 0.05$\le$x$_{Ca}$$<$0.125 range where two types of spin coupling are clearly identified,
associated with two spin wave branches separated by two anisotropy gaps\cite{Biotteau}.
This has been related to a charge segregation picture, indicated by the existence of ferromagnetic platelets\cite{Hennion1}.

In the present case, the coexistence of at least two dispersions at low q also suggests a charge segregation picture, in agreement with predictions\cite{Yunoki}, corresponding to two ferromagnetic states. By continuity with the low doping case, the quadratic spin wave branch defined for {\bf q}$<${\bf q$_0$} (long distance coupling), could be characteristic of hole-rich regions, and the "MV" type of coupling, specially observed on a local scale, to hole-poor regions with mainly localized electrons. The isotropy of the low-q dispersion curve, suggests a liquid orbital state as found in the metallic state\cite{Nagaosa}. 
The disappearance of the SE type of coupling, clearly identified from the anisotropy of the higher energy branch for x$<$0.125\cite{Biotteau},
% replaced by the new MV coupling for q$>$q$_0$, 
could reveal the effect of orbital degeneracy below this low-temperature transition.
These two ferromagnetic media, likely already exist dynamically, below T$_C$, in the metallic state, as indicated by broad magnetic modes. They become stabilized below 90K as they are involved into a collective state corresponding to their periodic arrangement, revealed by the superstructure. This transition corresponds to a {\it step-increase} of the magnetic excitations in agreement with the small increase of the magnetization.

 The similarities found for the spin dynamics of the x$_{Sr}$=1/8\cite{Moussa} and x$_{Ca}$=0.2 samples suggest that both samples are in a close magnetic state, specific of the 1/8 stoichiometry. In the x$_{Ca}$=0.2 sample, the extra-holes could be trapped as polarons giving rise to the observed diffuse scattering\cite{Dai1}, playing no r\^ole in this transport-magneto-structural transition, although likely responsible for the weak intensity of the superlattice peak and its metastability. 
The simplest picture would consist of hole-rich layers, stacked along [001], with a 4a$_0$ periodicity according to the model of Inami reported in Ref. 2, in continuity with our 2D-platelets picture. However, the disagreement of this picture with a quantitative analysis of the diffraction pattern, and the observation
 of the {\it isotropy} of the {\bf q$_0$} anomaly in the spin dynamics, requires a more complex arrangement\cite{Yamada2}.

This analysis of the spin dynamics, in terms of two distinct types of coupling, associated with distinct anisotropy gaps, agrees with the mixed magnetic state derived from nuclear magnetic resonance study of La and Mn atoms\cite{Papavassiliou,Belesi}. 
 We outline that a magnon-phonon coupling has been invoked in the metallic state at larger doping, where a locking of magnons on a flat phonon branch has been observed\cite{Dai2}. The present case differs from the metallic one, since the dispersion of the magnetic excitations
does not soften at large q, but rather, approximately follows a cosine-law as for usual insulating ferromagnets\cite{Hennion1}. 
The indirect coupling of magnons with Jahn-Teller phonons through orbital fluctuations previously proposed\cite{Khaliullin} is unsufficient here, where acoustical phonons are involved. Also, the usual magnon-phonon coupling\cite{Furukawa} cannot explain the resonance effect observed here. Whereas elastic forces are invoked to explain stripe or sheet structures for charge distribution\cite{Khomskii}, this spin dynamics reveals the r\^ole of new magneto-elastic forces to stabilize a collective state of two ferromagnetic media, which remains to be determined.


\begin{thebibliography}{30}

\bibitem{Yamada}Y. Yamada et al,
Phys. Rev. Lett. {\bf 77}, 904 (1996)
\bibitem{Yamada2}Y. Yamada et al.
 Phys. Rev B {\bf 62}, 11600 (2000)
\bibitem{Pinsard}Pinsard et al.  Physica B {\bf 234-236}, 856 (1997)

\bibitem{Endoh}Y. Endoh et al. 
 Phys. Rev. Lett. {\bf 82}, 4328 (1999)
\bibitem{Okuda} T. Okuda et al. Phys. Rev B {\bf 61} 8009 (2000)
\bibitem{Biotteau} G. Biotteau et al. Phys. Rev. B {\bf 64} 104421 (2001)
\bibitem{Moussa} F. Moussa et al. ICNS  (2001) to appear in J. of App. Phys.
\bibitem{Doloc}L. Vasiliu-Doloc et al. Phys. Rev B {\bf 58}, 14913 (1998)

\bibitem{Lovesey}S. W. Lovesey "Theory of neutron scattering from Condensed mater" Vol. 2 Clarendon Press Oxford (1972) 
\bibitem{Hennion} M. Hennion et al. Phys. Rev B {\bf 61} 9513 (2000)

\bibitem{Yunoki} S. Yunoki  et al. Phys. Rev. Lett. {\bf 81}, 5614 (1998)
\bibitem{Nagaosa}S.Ishiara et al.Phys. rev B {8bf 56}, 686 (1997)
\bibitem{Dai1}Pencheng Dai et al. Phys. Rev. Lett. {\bf 85}, 2553 (2000)
\bibitem{Hennion1}M. Hennion et al. SCES (2001) to appear in Physica B.
\bibitem{Papavassiliou} G. Papavassiliou et al. Phys. Rev. Lett.{\bf 84}, 761 (2000)
\bibitem{Belesi} M. Belesi et al. Phys. Rev. B {\bf 63}, 180406R (2001)
\bibitem{Dai2} Pengcheng Dai et al. Phys. Rev B {\bf 61} 9553 (2000)
\bibitem{Khaliullin}G. Khaliullin and R. Kilian Phys. Rev. B {\bf 61} , 3494 (2000)
\bibitem{Furukawa}N. Furukawa J. Phys. Soc. Jpn {\bf 68} 2622 (1999)
\bibitem{Khomskii}D. I. Khomskii and K. I. Kugel Eur. Phys. Lett. {\bf 55}, 208 (2001)

\end{thebibliography}
\end{document}